\documentstyle[11pt]{article}
\input{psfig}
\begin{document}
\title{
Effects of magnetohydrodynamics matter density fluctuations on 
the solar neutrino  resonant spin-flavor precession}
\author{N. Reggiani$^{a,1}$, M.M. Guzzo$^{b,2}$, J.H. Colonia$^{b,3}$ and \\  P.C. de Holanda$^{b,c,4}$
\\
\vspace{0.1cm}
$^a$ {\it Instituto de Ci\^encias Exatas} \\
{\it Pontif\'\i cia Universidade Cat\'olica de Campinas} \\
{\it 13020-970 Campinas, SP, Brasil} 
\\ 
\vspace{0.1cm}
$^b${\it Instituto de F\'\i sica Gleb Wataghin} \\
{\it Universidade Estadual de Campinas - UNICAMP} \\
{\it  13083-970 Campinas, S\~ao Paulo, Brasil} 
\\ 
\vspace{0.1cm}
$^c$ {\it Instituto de F\'\i sica Corpuscular} \\
{\it Universitat de Val\`encia} \\
{\it 46100 Burjassot, Val\`encia, Spain}
}
\date{}
\maketitle

\begin{abstract}

Taking into account the stringent limits from helioseismology observations
on  possible matter density fluctuations described by magnetohydrodynamics
theory, we find the corresponding time variations of solar neutrino survival
probability due to the resonant spin-flavor precession phenomenon with
amplitude of order $O(10$\%). We discuss the physics potential of high
statistics real time experiments, like as Superkamiokande, to observe
the effects of such magnetohydrodynamics fluctuations on their data. We
conclude that these observations could be thought as a test of  the resonant 
spin-flavor precession solution  to  the solar
neutrino anomaly.
\end{abstract}
 
\vskip.1cm



\newpage 

\section{Introduction}

According to all attempts to build a solution to the solar
neutrino anomaly based on the resonant spin-flavor precession
mechanism~\cite{rsfp}, when one assumes a neutrino magnetic moment of
order $\mu_\nu=10^{-11}\mu_B$, which is slightly below its present
experimental limit~\cite{pdg}, an average magnetic field of order
$10^{4}$~G~\cite{alp,gn98} is required  to conciliate the present
solar neutrino experimental data~\cite{homestake}-\cite{superkamioka} and
solar neutrino theoretical predictions~\cite{JBahcall}-\cite{BKS}.  Using
these typical values for the solar magnetic field we observed~\cite{GRC97}
that magnetohydrodynamics~\cite{priest,foukal} naturally predicts the existence of solar
magnetic oscillations which period is of the order of 1 to 10 days.
Resonant spin-flavor precession mechanism generates a
solution to the solar neutrino anomaly only if neutrinos are very
sensitive to the magnetic field as well as matter density profiles
inside the Sun and, in particular, are very sensitive also to the
magnetohydrodynamics fluctuations of these quantities.  Therefore we
expect that these fluctuations will generate an accordingly fluctuating
experimental neutrino counting rate. The corresponding observable
signal of this phenomenon is a periodical time fluctuation of the
solar neutrino flux, in particular, in the high energy portion
of the solar neutrino spectrum, detected in present real time
experiments~\cite{superkamioka,sno}.  
These effects can be thought as a test to the resonant spin-flavor
precession solution to the solar neutrino problem since they can clearly
be distinguished from other sources of time modulation in the solar
neutrino observations like the effect of seasonal variation of the
Sun-Earth distance on vacuum oscillations, MSW effects at the Earth,
and solar magnetic activity related to the appearance of sunspots on
the solar surface.

In general, magnetohydrodynamics~\cite{priest,foukal} predicts two 
different kinds of stable
fluctuations. Alfv\'en waves, when transverse oscillations of the
fluid with respect to the magnetic field and the propagation vector are
present and, therefore, no fluctuation of matter density or pressure
is observed. And magnetosonic waves, which present longitudinal
oscillations of density with respect to the propagation vector.
In reference~\cite{GRC97} we analyzed the effects of purely Alfv\'en
waves where only magnetic fluctuations affects the solar neutrino
survival probability. Nevertheless, if matter density perturbations are
present in the way described by magnetohydrodynamics, their effects can
be important for the relevant high energy neutrinos observed in real
time experiments~\cite{superkamioka,sno}.
There are two main reasons which can
be evoked to justify this. The presence of matter suppresses, in general,
the  neutrino spin-flavor precession mechanism, once that matter appears
in the diagonal entries of the evolution matrix, which are spin-flavor
conserving. That is why the presence of a resonance, a position along
the neutrino trajectory where the matter effects are cancelled  by a term
proportional to the squared mass  difference divided by the neutrino
energy, is important for the neutrino spin-flavor conversion. Therefore,
fluctuations of the matter density can generate modifications in this
resonance  cancellation leading to important variations on the final
probability of neutrino spin-flavor precession. The second reason to 
consider the effect of density fluctuations is 
the peculiar behavior of the adiabaticity parameter in the resonant
spin-flavor precession phenomenon. In fact, several authors discussed
the effect of density variations on the solar neutrino matter enhanced
transitions, the MSW effect~\cite{density}. In this context, they
conclude that the effects of density perturbations on transitions of
solar neutrinos may be substantial if these perturbations occur in the
central region of the Sun and therefore only low energy neutrinos are
sensitive to such effects since their resonance layer occurs deeper in
the Sun. Note however that this conclusion cannot be straightforward
invoked here. While the adiabaticity increases with increasing resonance
density for the MSW effect, it is inversely proportional to the resonance
density for the spin-flavor precession phenomenon and the effects of
density perturbations in the solar convective zone may be important. This
low density region is relevant for high energy solar neutrinos since
a resonance for such neutrinos can be found there. Consequently high
neutrinos observed in real time experiments can be sensitive to matter
fluctuations in this region.

In this paper we investigate the impact of  matter density fluctuations
induced by sonic magnetohydrodynamics waves on observable time fluctuations
of high energy solar neutrino flux detected in real time experiments and
compare these effects with consequences from  purely magnetic (Alfv\'en)
fluctuations.  We conclude that both magnetic and density fluctuations 
can be equally relevant,
although matter density fluctuations are much more constrained from
helioseismology observations. Taking into consideration such constraints,
we still expect to find periodical time fluctuations in solar neutrino
counting rate due to magnetohydrodynamics matter density waves of order
of 10\% of the total number of events.

\section{Magnetohydrodynamics and neutrino spin-flavor precession}

We are assuming a non-vanishing neutrino transition magnetic moment.
In this case, the interaction of neutrinos with a magnetic field will
generate spin-flavor precession which is given by the evolution
equations \cite{pulido}

\begin{eqnarray} & \hskip-.8cm i{\displaystyle{ d }\over 
\displaystyle{ dr}}\left( 
\begin{array}{c} \nu_R
\\ \nu_L \end{array} \right) = 
\left( \begin{array}{cc} {\sqrt{2}\over 2}G_F\alpha N_e(r) -
{\Delta\over 4E} & \mu_\nu |\vec B_\perp(r)|
 \\ \mu_\nu |\vec B_\perp(r)| & -{\sqrt{2}\over 2}G_F\alpha N_e(r) +
{\Delta\over 4E}  \end{array} \right)
\left( \begin{array}{c} \nu_R \\ \nu_L
\end{array} \right) \label{motion} \end{eqnarray}

\noindent where $\nu_L$ ($\nu_R$) is the left (right) handed component of
the neutrino field, $\Delta= |m_L^2-m_R^2|$ is their squared mass
difference, $E$ is the neutrino energy, $G_F$ is the Fermi constant,
$N_e(r)$ is the electron number density distribution and $|\vec
B_\perp(r)|$ is the transverse component of the  magnetic field.  
Finally we have $\alpha=5/6$ for Majorana
neutrinos, in which case the final right-handed states $\nu_R$ are active
nonelectron antineutrinos.  For Dirac neutrinos, $\alpha=11/12$, in which
case the right-handed final states are sterile nonelectron
neutrinos~\cite{pulido}.  In this paper we will assume Majorana neutrinos. 
Note however that the multiplicative factor 11/10 which has to be included
in $\alpha$ for Dirac neutrinos does not lead to important alterations in
our conclusions, which are, in this way, valid for Majorana or Dirac
neutrinos. 

From Eq.~(\ref{motion}) we observe that the neutrino spin-flavor
precession mechanism depends on both the magnetic field  $|\vec
B_\perp(r)|$ as well as the matter density distribution $N_e(r)$ along the solar
neutrino trajectory. These quantities are affected by magnetohydrodynamics
fluctuations. In order to describe them,  we consider the linearized
magnetohydrodynamics equations~\cite{priest,foukal}, following the same
steps of reference~\cite{GRC97}, where magnetohydrodynamics spectrum
is generated by small displacements {$\vec \xi$} from an equilibrium
configuration.  We are particularly interested in the magnetic and
matter density fluctuations generated by the displacement $\vec \xi$
given, respectively, by~\cite{goedbloed} \begin{equation} \delta{\vec
B}={\vec\nabla}\times({\vec\xi} \times {\vec B_0}) \hskip1cm {\rm and}
\hskip1cm \delta\rho=\vec\nabla \cdot (\rho\vec\xi).
\label{pert_mag} 
\end{equation}
where $\vec B_0$ is the equilibrium magnetic field configuration
assumed to be the one relevant to the solar neutrino problem described
in reference~\cite{alp} and $\rho$ is the solar matter distribution
calculated in the standard solar model~\cite{JBahcall}-\cite{INT}. 
 
A crucial region of the magnetohydrodynamical spectrum is the continuum
region which is associated with singularities of the Hain-L\"ust 
equation~\cite{GRC97,goedbloed}.
This happens when the frequency of the magnetohydrodynamical fluctuation
is equal to
\begin{eqnarray}
w_A^2 &=& {{k^2B_0^2}\over{\rho}}\hskip1cm {\rm or} \hskip1cm w_S^2 = 
{{\gamma p}
\over{\gamma p + B_0^2}}{{k^2B_{0}^2}\over {\rho}}, 
\label{continua}
\end{eqnarray}
where $p$ is
the pressure and $\gamma=C_p/C_v$ is the ratio of specific heats.
$w^2=w_A^2$ and $w^2=w_S$ define the Alfv\'en and slow continuum 
regions in the 
magnetohydrodynamical spectrum~\cite{goedbloed}. Magnetic 
waves and matter
density fluctuations associated with these frequencies are called
localized modes since they present the interesting feature of
been highly peaked around the position, $r_s$,  where the singularity
occurs~\cite{accumulation}.

In order to numerically overcome the singularities associated with the
continuum spectra, a resistivity layer in the position of the 
singularity is
introduced~\cite{villard}. The width of this layer is directly
related with the width $\delta r$ of the localized magnetic or 
matter density
fluctuation which can be estimated~\cite{tataronis}:
\begin{equation}
\delta r \approx 8 \pi
\left( {\eta \over {\mu_0 \omega(r_s)}} \right)^{1/3}
\left( {{2 B'} \over B} - {\rho' \over \rho} \right)^{-1/3},
\label{width}
\end{equation}
where $\mu_0$ is the vacuum permeability. In our particular 
scenario, $\delta r$ 
can achieve 10$^{-1}$ (normalized by the solar radius).  

To analyze the consequences of the localized waves on the solar neutrino
observations, we have now to define the amplitude of possible magnetic
and matter density fluctuations. The solar magnetic
field is relatively free to fluctuate since the magnetic pressure
$B^2/8\pi$  is negligibly small when compared with the dominant
gas pressure $p$~\cite{JBahcall} if we consider the matter 
density distribution $\rho$
predicted by the standard solar model and the  magnetic field
strength of order of those ones required to solve the solar neutrino
anomaly~\cite{alp}. 
In fact, in this case, $B^2/8\pi p$ varies from approximately $10^{-6}$
in the central regions of the Sun to order $10^{-4}$ close to the solar
surface. From this argument,  the magnetic field can be as large as 10$^9$
G in the solar core or   10$^7$ G  in the solar convective zone.
More stringent bounds on the magnetic field in the convective zone 
are found in refs. \cite{SR,shi} where the discussion is 
based on the non-linear effects which eventually prevents the growth of 
magnetic fields created by the dynamo process. 
By equating the magnetic tension to the 
energy excess of a sinking element at the bottom of the
convective zone, Schmitt and Rosner \cite{SR} 
obtained few times $10^4$~G as an upper bound.
Therefore fluctuations of the solar magnetic
field of  the same order of magnitude of the magnetic field
used in reference~\cite{alp} can be found inside the Sun.
Despite this fact, they cannot be arbitrarily large when we are considering
the linearized magnetohydrodynamics equations~\cite{GRC97,goedbloed}. This
implies that the displacements $\vec \xi$ must be small, $|\vec \xi | <1$, such
that the non-linear terms can be neglected. This implies that $|\delta \vec
B|/|\vec B_0| < 1$. The error associated with this approximation is
$\sigma\approx  (|\delta \vec B|/|\vec B_0|)^2$. The maximum possible value
for the ratio $|\delta \vec B|/|\vec B_0|$ is related with a clear statistical
distinction between the maximum and the minimum value of the perturbed
magnetic field, which is given approximately by $(|\delta \vec B|/|\vec
B_0|)/\sigma$ (in units of $\sigma$). To have a minimum 2-$\sigma$
distinction between the maximum and minimum magnetic field, we must have a
maximum value of the perturbation $\approx |\delta \vec B|/|\vec
B_0| = 0.5$.

Something very different happens with possible matter density
fluctuations. In fact, they are very constrained by helioseismology
observations. The largest density fluctuations
$\delta \rho$   inside the Sun are induced by temperature fluctuations
$\delta T$ due to convection of matter between layers with different
local temperatures. An estimate of such effect is presented in
reference~\cite{hiroshi_npb96} and gives
\begin{equation}
{\delta \rho\over \rho} = m_p g(r-r_0){\delta T \over T^2} = {r-r_0\over R_0} 
{\delta T \over T}
\label{amplitude}
\end{equation}
where $m_p$  is the nucleon mass, $g(r)$ is the gravity acceleration and
$R_0\approx 0.09 \times R_\odot$ ($R_\odot$ is the solar radius) is a
numerical factor coming from the approximately exponentially decreasing
standard matter density distribution inside the Sun~\cite{JBahcall}. Since
${\sqrt{<\delta T^2>}}/T \approx 0.05$ is not in conflict with
helioseismology observations~\cite{9dohiroshi}, taking $(r-r_0)/
R_0\approx 1$, we have to consider density fluctuations $\delta \rho/\rho$
smaller than 10\%. In fact, in an accurate analysis of helioseismology
consequences on matter density fluctuations~\cite{fiorentini} it was
concluded $\delta \rho/\rho$ can be very large (larger than 10\%) only for
very inner parts of the Sun ($r<0.04$) as well as for very superficial
regions ($r>0.98$). For $0.04<r<0.25$, $\delta \rho/\rho$ decreases
approximately linearly and achieves its smaller value 2\% in $r\approx
0.25$. Finally in the region where $0.4<r<0.9$, $\delta \rho/\rho$ is
approximately 5\%.  We impose these constraints as boundary conditions
for the amplitudes of density fluctuations we will consider in the following.

Finally we can analyze the consequences of the localized
magnetohydrodynamics modes on the solar neutrino flux
solving the neutrino evolution equations when a non-vanishing neutrino
transition magnetic moment $\mu_\nu$ is assumed. We will adopt
here a phenomenological approach, in close analogy to what was done
in reference~\cite{GRC97}. We will assume   that magnetohydrodynamics
introduces gaussian shaped fluctuations which will be added to the
equilibrium configuration of both matter density and/or magnetic field
profile. This gaussian perturbation is centered in $r_s$, with width
$\delta r$. For the matter density fluctuation we have:

\begin{equation}
 \delta \rho(r,t) = \epsilon_\rho \rho(r_s)\exp \left[ -\left({r-r_s\over 
\delta r} \right)^2 \right] \sin\left[ w(r_s)t \right], 
\label{rhopert}
\end{equation}
where $\epsilon_\rho$ is the fluctuation amplitude normalized to
the value of the standard matter density $\rho(r)$ calculated in the
position of the singularity $r_s$ and $\delta r$ is the width of the 
fluctuation. The frequencies $w(r_s)=w_A$ or $w(r_s)=w_S$ are given
in equation (\ref{continua}) and introduce a periodical time modulation
on the standard matter density profile.  Similarly, our assumption for
the magnetic fluctuations is obtained from the above equation substituting
$\rho\to B_0$, i.e., the standard matter density by the equilibrium field
profile.
Therefore the final matter density (as well as the magnetic
field profile) will be given by some equilibrium profile summed to the
perturbation shown in Eq.~(\ref{rhopert}). 
We will assume the parameter
$\epsilon_\rho$ and $\epsilon_B$ varying from 0 to 0.05, which means
that the matter density as well as the magnetic field fluctuate around
their equilibrium values with maximum amplitude around 5\% of this
value.
These fluctuations  will generate an according 
time fluctuation of the neutrino counting governed by the evolution 
equations (\ref{motion}).

Since we have  a very stringent experimental limit on the neutrino
magnetic moment~\cite{pdg}, a large magnetic field is necessary to find
a relevant spin-flavor conversion of neutrinos which propagating through
it. In this paper we will consider a magnetic field profile proposed in 
reference \cite{alp} which is as large as $10^6$ to $10^7 G$
in the central regions of the Sun and fall by  two orders of
magnitude when the convective zone is reached~\cite{field}:
\begin{equation} B_{0}(r)=\left\{
\begin{array}{lcl} 
 a_1 \left({0.2\over r+0.2}\right)^2 {\rm G} & {\rm for} & 0 < r \leq 0.7
\\ 
B_C & {\rm for} & r > 0.7  ~ ,
\end{array} 
\right.  \label{Akhmedov} \end{equation}
where $B_C$ is the magnetic field in the convective zone given by the 
following
profiles:
\begin{equation}
B_C=a_2 \left[1-\left({r-0.7 \over 0.3}\right)^n\right] {\rm G} ~~~ 
{\rm for} ~~ 
0.7< r \leq 1.0 
\label{prof1}
\end{equation}
or
\begin{equation}
B_C=a_2\left[ 1+\exp\left({r-0.95 \over 0.01}\right) \right]^{-1}  
{\rm G} ~~~ 
{\rm for} ~~ 0.7< r \leq 1.0 ~ .
\label{prof2}
\end{equation}
$a_1\approx 10^5-10^7$ and $a_2\approx 10^4 - 10^5$ in such a way that the
continuity of the magnetic field at the point $r=0.7$ is satisfied and
$n=$ 2, 6 and 8.  We assume also that the magnetic equilibrium profile
$B_0(r)$ is in the $z$ direction. For the solar scenario, $\gamma p >>
B_0^2$ and therefore $w_A\approx w_S$. 

In Figure~1 we show the periods of the magnetic and matter density
fluctuations associated with the continuum spectra given in
(\ref{continua}) for several  magnetic field profiles given by the Eqs. 
(\ref{Akhmedov})-(\ref{prof2}).
We observe that for the considered magnetic fields 
typical periods vary from $O(1)$ to $O(10)$
days. Time fluctuations of solar neutrino observations presenting periods
of this order of magnitude  constitute the most important signal of the
existence of solar magnetohydrodynamics fluctuations and  a crucial test
for the resonant spin-flavor neutrino precession mechanism to the solar
neutrino problem~\cite{footnote}.

\section{Results}

The results of our analysis are shown in Figure 2. The fluctuations of
the magnetic and matter density induce fluctuations in the survival
probability $P(\nu_L \to \nu_L)$, which is calculated solving
the Eqs.~(\ref{motion}). In this figure we present the amplitude
$\Delta P$ of the fluctuations of the survival probability as a  function
of $\Delta m/4E$, for some values of $r_S$, the position
of the localized mode.  We show the phenomenon for $r_S=0.5$, 0.9 and
for the value of $r_S$ which gives the maximum probability amplitude,
usually around $r_S\approx 0.7$. We adopted the magnetic field profile
given in Eqs.~(\ref{Akhmedov}) and (\ref{prof1}), with $n=6$ and $\mu_\nu=1\times 10^{-11}\mu_B$. 
 Other magnetic field profiles
shown in Eqs.~(\ref{Akhmedov}), (\ref{prof1}) and (\ref{prof2})  generate  
very similar
consequences of magnetohydrodynamics fluctuations on the solar neutrino
survival probability and we will not show them here.  We assumed
$\epsilon_\rho$ and $\epsilon_B$ in Eq.~(\ref{rhopert}), i.e., the
relative amplitude of matter density and magnetic field profile
fluctuations, respectively, to  vary from 0 to 5\% as it is indicated
in  Figure 2. Also, vertical lines indicate the value of Log$(\Delta
m/4E)$ corresponding to a resonance coinciding with the indicated
singularity $r_S$. We can observe that the fluctuations of the survival
probability are maxima near the regions of the resonances and are well
localized in $\Delta m/4E$ for $r_s \le 0.7$.

We verified also  that, for fixed values of $r_S$ and   $\Delta m/4E$, the
corresponding probability amplitude varies linearly with the magnitude of
the amplitude of the matter density $\delta\rho$ and/or magnetic field
fluctuation $\delta \vec B$.  For instance, when $\epsilon_\rho=\epsilon_b$, the
maximal probability fluctuation corresponding to the magnetic fields
given in Eq.~(\ref{prof1}) occurs in $r_S=0.69$ and can be written as
$\Delta P \approx   2.20\times 10^{-2} \epsilon_\rho$ for $n=2$, $\Delta
P \approx   2.48\times 10^{-2} \epsilon_\rho$ for $n=6$ and $\Delta P
\approx   2.64\times 10^{-2} \epsilon_\rho$ for $n=8$.  Considering the
exponential behavior for the magnetic field in the convective zone,
the maximal amplitude of survival probability  occurs in $r_S=0.68$ and
presents the following linear behavior with  $\epsilon_\rho$: $\Delta
P \approx 2.81\times 10^{-2} \epsilon_\rho$. Therefore, given the results of
Figure 2, one can easily infer the amplitude of the survival probability
for other values of $\delta\rho$ and $\delta \vec B$.
 Note that this approximately linear
behavior is valid for $\epsilon_\rho<0.2$, which includes the physical
limit for helioseismologically acceptable variations of the matter density.

From Figure 2, we observe also that the maximal effect of the matter
density and/or magnetic field fluctuation on the survival probability
occurs if these magnetohydrodynamics fluctuations are found in the beginning
of the convective zone where $r_S\approx 0.7$. This can be understood when
we examine the behavior of the adiabaticity parameter for the spin-flavor
conversion phenomenon. As we have already mentioned, this parameter
increases with $r$ and approaches the unit for $r\approx 0.7$. In
fact, it was pointed out previously (see, for instance~\cite{GB92})
that this is a privileged situation for the neutrino  conversion.
Note also that according to references \cite{alp,gn98}, the required value
of $\Delta m$ to find a solution to the solar neutrino anomaly is
$O(10^{-7})$~eV$^2$ or smaller. If $\Delta m=O(10^{-7})$~eV$^2$,
neutrinos of $O(10)$~MeV in energy will experience their resonance
around $r=0.7$ and therefore will be very sensitive to the maximal
amplitude probability indicated in Figure~2.  This is exactly the energy
range probed by real time experiments,
like as Superkamiokande~\cite{superkamioka} and SNO~\cite{sno}, which
are, therefore, very suitable to investigate the hypothesis of solar
magnetohydrodynamics perturbations and their effects on solar neutrinos.

After several hundred days of taking data, Superkamiokande observations
are, in principle, able to verify the existence of time fluctuation with
period from $O(1)$ to $O(10)$  days. Unfortunately, these data are not
published nor available. Nevertheless we can argue the potential of
this experiment to perform such analysis. Figure~3 shows the amplitude
of the expected event rates in Superkamiokande experiment when we assume
that a magnetohydrodynamics fluctuation is localized around $r_S\approx
0.7$. We convulated the standard $^8B$-neutrino production
spectrum~\cite{JBahcall}, the $\nu_e - e^-$ scattering cross-section
increasing linearly with energy and the experimental efficiency which
is approximately 20\% for neutrino energies around 5 MeV up to a
maximum of 70\% for neutrino energies of 10 MeV or larger.  These event
rates are given in solar neutrino units (SNU)
 for several values of $\Delta m$. We observe that for $\Delta m = 3
 \times 10^{-7}$~eV$^2$, the maximum event rate amplitude is obtained.
 I.e., after having been precessed due to the existence of a sonic
 magnetohydrodynamical waves, the number solar neutrinos scattering
 with Superkamiokande electrons fluctuates with amplitude around
 0.003-0.004 SNU, if these neutrinos present energies around 8 and
 10~MeV. This is approximately 10~-~15\% of the total rate of
 scattering events in this energy range. Note that if $\Delta m$ is
 different from the values shown in Figure~3, this means that neutrinos
 with energies around 5~-~15~MeV will experience resonance in a
 position different from $r_s\approx 0.7$ and, although not maximal,
 their counting rate will still fluctuate.

\section{Conclusions}

 Magnetohydrodynamics predicts magnetic as well as  matter
density fluctuations in the Sun. The
purpose of this paper is to analyze the effects of possible solar
matter density fluctuations on solar neutrino observations taking into
account the stringent limits on these fluctuations coming from
helioseismology. We  verify also the potential of high statistics real
time solar neutrino experiments, like as Superkamiokande, to observe
these effects. We showed that  the survival probability fluctuations
of active solar neutrinos due to the resonant neutrino spin-flavor
precession can be of order 10\% if amplitudes of the magnetic or matter
density fluctuations $\epsilon_b$ or $\epsilon_\rho$ are of order of 5\% (20\% if
$\epsilon_b=\epsilon_\rho=$10\%). Therefore, in order to make evident  the existence of
these magnetohydrodynamics fluctuations in the Sun, we have to look for
solar neutrino data time fluctuations of this order in the time scale
compatible with the periods predicted by the theory. From Figure~1, we
see that relevant time scale can be  of order $O(1)$ to  $O(10)$ days. Collecting
$O(20)$ events a day, Superkamiokande presents a statistics error of
approximately 10\% in an interval of 5 days. Furthermore, a Fourier
analysis of the experimental observations fitting a period of time of
the order of the total period of Superkamiokande observation (of order
of hundred days), could drastically reduce the involved errors and the
amplitudes of order of 10\% shown in Figures 1 and 2 could be
experimentally tested.  There is no  available published solar neutrino
data to conduct a conclusive analysis of possible data fluctuation with
period $O(1-10)$ days. Analysis of time fluctuations on solar neutrino
data conducted up to now, privileged different periods: exact 24 hours
as a result of MSW effect inside the Earth, six months to conciliate
seasonal variation and neutrino vacuum oscillations or 11 years to
verify solar magnetic activity effects on spin flip phenomenon.

We conclude that Superkamiokande is potentially interesting to
investigating the effects on solar neutrino observations coming from
magnetohydrodynamics fluctuations.  The observation of these effects
could be taken as an evidence of the resonant spin-flavor precession
mechanism in the Sun.

\vspace{1cm}

{\large \bf Acknowledgements}: 
This work was supported by Funda\c c\~ao de Amparo \`a Pesquisa do Estado
de S\~ao Paulo (FAPESP), Programa de Apoio a N\'ucleos de Excel\^encia
(PRONEX) and Con\-se\-lho Na\-cio\-nal de De\-sen\-vol\-vi\-men\-to
Cien\-t\'\i \-fi\-co e Tec\-no\-l\'o\-gi\-co (CNPq).

\hrule

$^1$ Email: ice-fisica@acad.puccamp.br

$^2$ Email: guzzo@ifi.unicamp.br

$^3$ Email: colonia@ifi.unicamp.br

$^4$ Email: holanda@ifi.unicamp.br

\newpage

\topmargin = 3cm

\begin{figure}
\psfig{figure=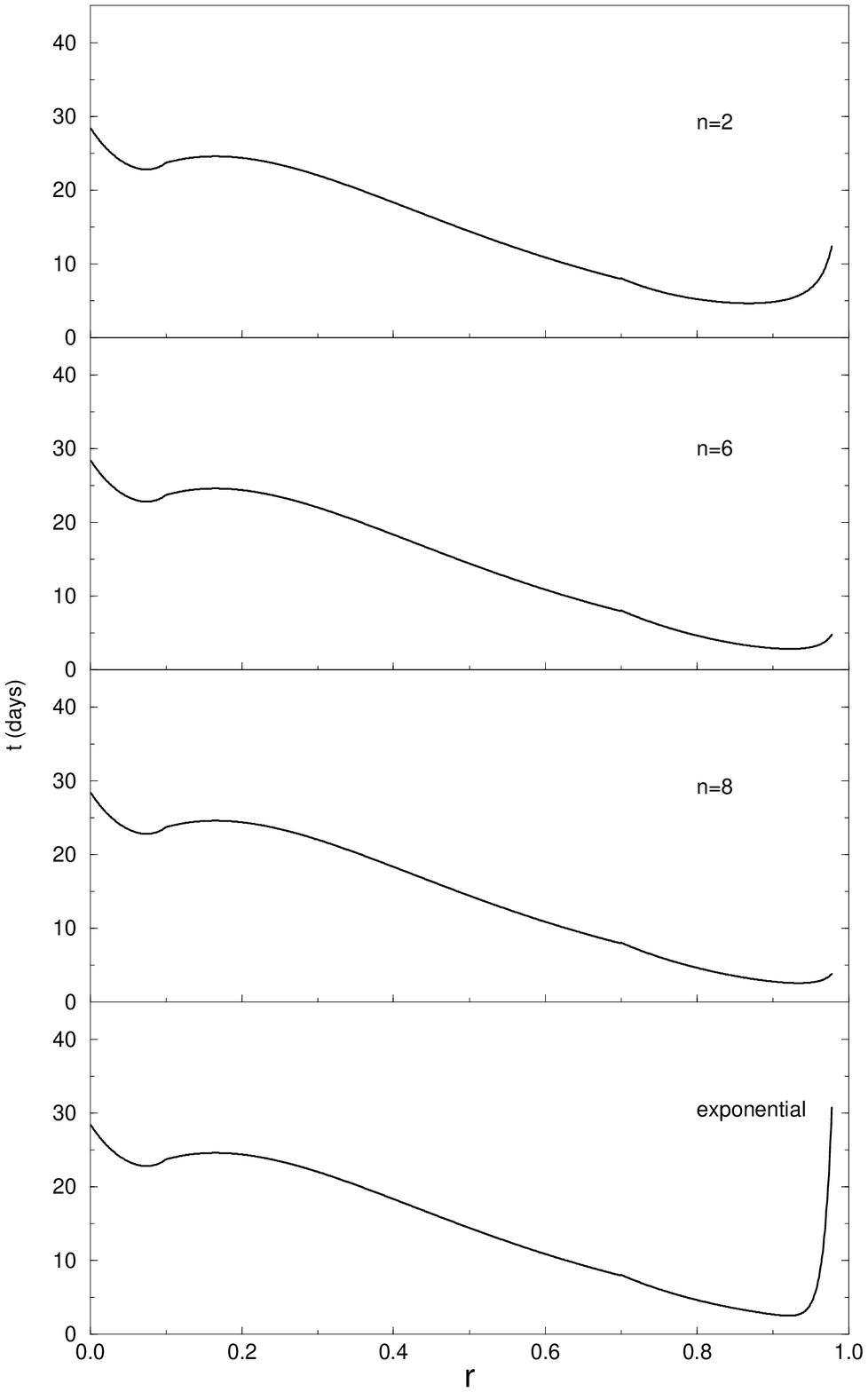,height=12cm} 
\end{figure}

{\bf Figure 1}: Periods of the magnetohydrodynamics fluctuations for
the continuum spectra when the magnetic profile shown in
Eq.~(\ref{prof1}) when $n=2$, $n=6$, $n=8$ and exponential
profile of Eq.~(\ref{prof2}). We adopted the parameters  $a_{1}\approx
1.0 \times 10^{6}$ G, $a_{2}\approx 4.0 \times 10^{4}$ G and $k\approx
10^{-10}$~cm$^{-1}$.0

\newpage
\topmargin = 6cm

\begin{figure}
\begin{center}
\psfig{figure=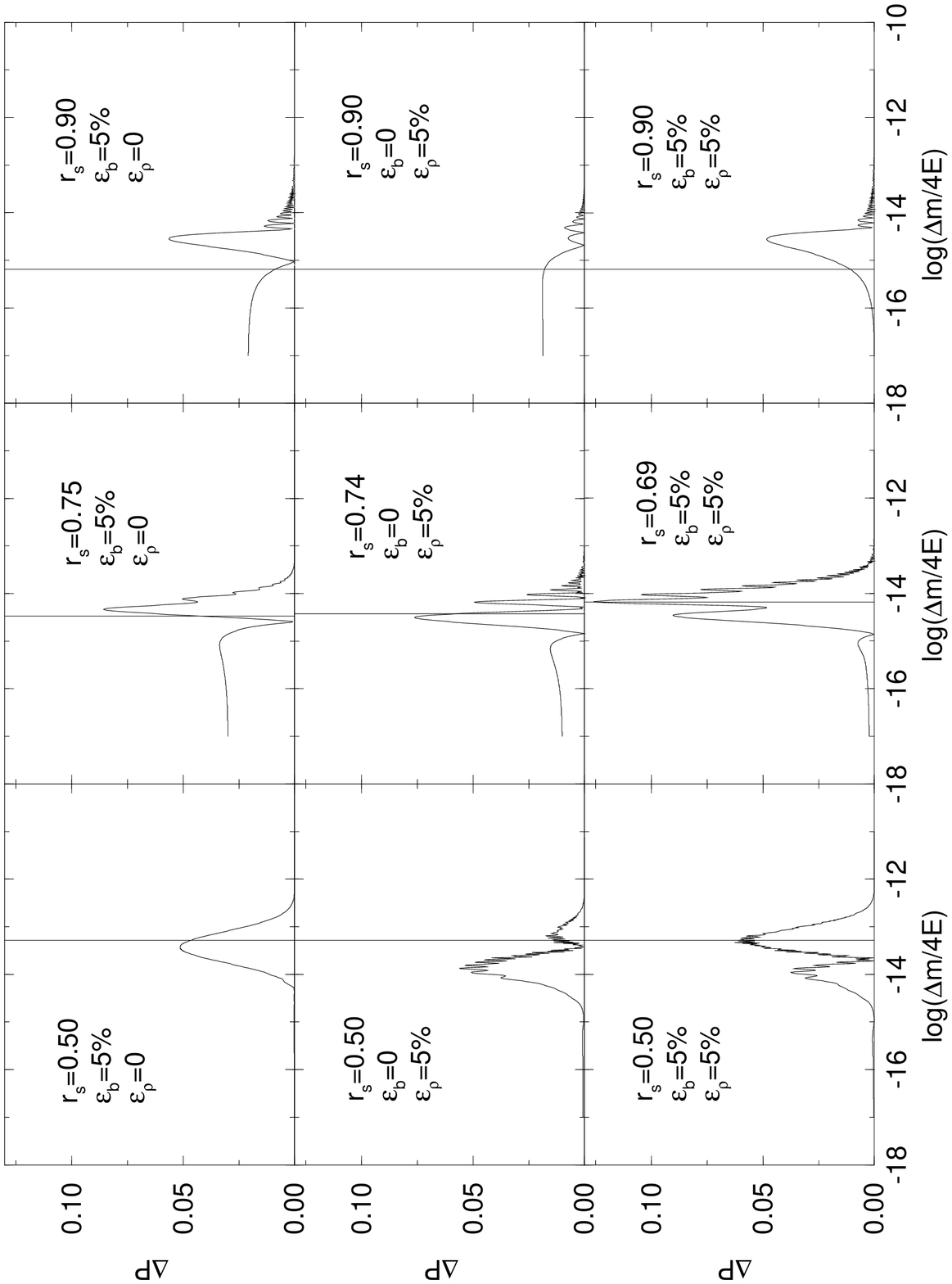,height=10cm}
\end{center} 
\end{figure}

\rule{0cm}{0.01cm}

\vspace{1.5cm}

{\bf Figure 2}: Amplitude $\Delta$P of the survival probability in
function of $\Delta m/4E$ for some values of $r_{s}$. The maximum
probability amplitude occurs around $r_{s}=0.7$.  The vertical lines
correspond to a resonance coinciding with the singularity $r_{s}$.

\newpage
\topmargin = 6cm

\begin{figure}
\psfig{figure=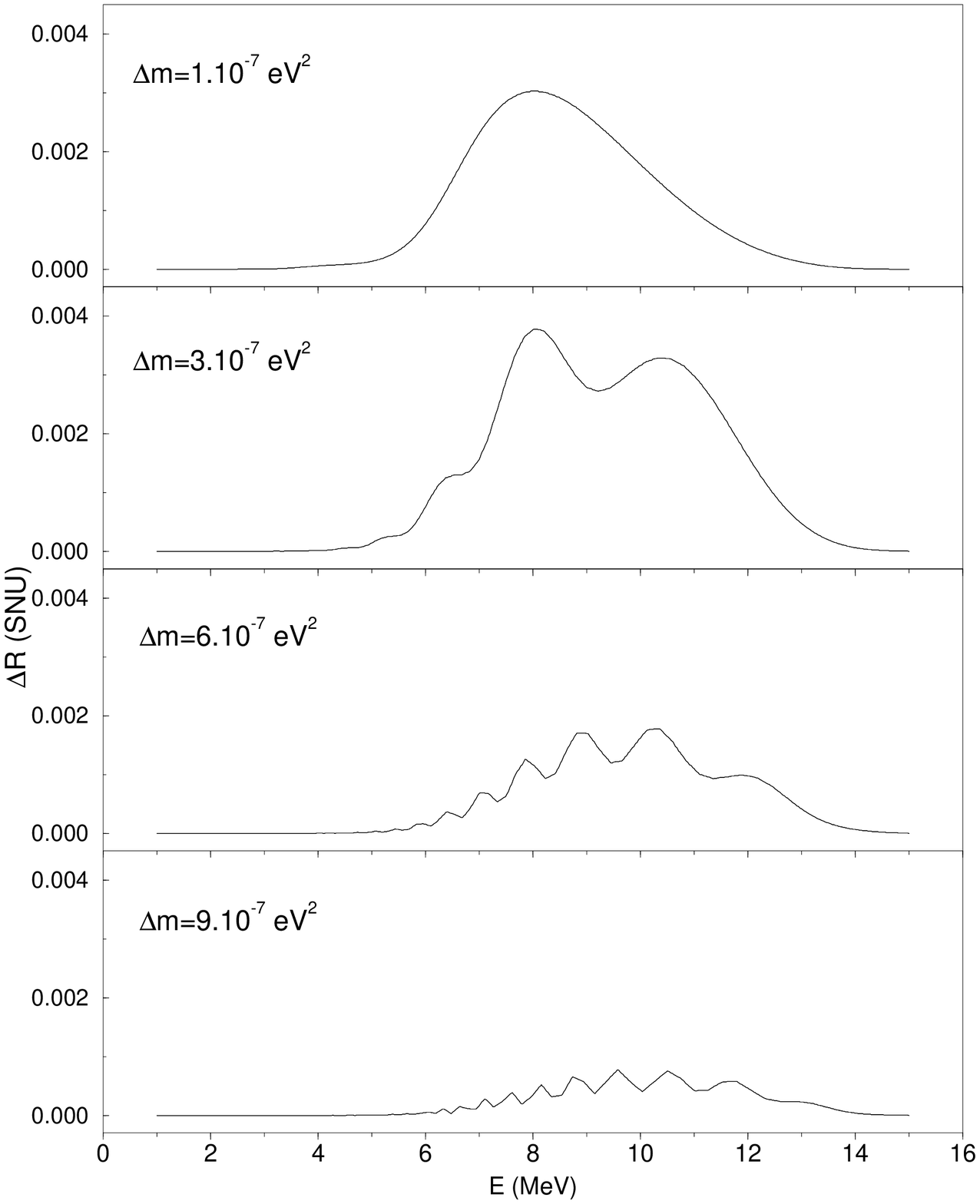,height=8cm} 
\end{figure}

{\bf Figure 3}: Amplitude $\Delta$R of the expected event rates to
Superkamiokande expe\-riment for some values of the mass square
difference $\Delta m$.


\begin{thebibliography}{99}

\bibitem{rsfp} A.  Cisneros, Astrophys.  Space Sci.
10 (1970) 87; M.  Voloshin, M.I.  Vysotsky and L.  Okun,
Sov.  J.  Nucl.  Phys.  44 (1986) 440; J.  Schechter and 
J.W.F.  Valle, Phys.
Rev.  D 24 (1981) 1883; E.  Akhmedov, Sov.  J.  Nucl.  Phys.
48 (1988) 382; Phys.  Lett.  B213 (1988) 64; C.S.  Lim 
and W.J.  Marciano, Phys.
 Rev.  D37 (1988) 1368.

\bibitem{pdg} C. Caso et al., ``Review of Particle 
Physics'', The European
Physical Journal C3 (1998) 1.

\bibitem{alp}E.Kh.  Akhmedov, A.  Lanza and S.T.  Petcov,
Phys.  Lett.  B 303 (1993) 85.

\bibitem{gn98} M.M. Guzzo and H. Nunokawa, ``Current status of the 
Resonant Spin-flavor Solution to the solar neutrino problem'', 
HEP-PH/9810408.

\bibitem{homestake}
K.\ Lande (Homestake Collaboration) in 
{\it Neutrino '98}, Proceedings of the XVIII International Conference on
Neutrino Physics and Astrophysics, Takayama, Japan, 4--9 June 1998,
edited by Y. Suzuki and Y. Totsuka, 
to be published in Nucl. Phys. B (Proc. Suppl.), 
scanned transparencies are available at 
URL http://www-sk.icrr.u-tokyo.ac.jp/nu98/scan/index.html.
 
\bibitem{kamiokande}
Y. Fukuda {\sl et al.} (Kamiokande Collaboration), 
Phys. Rev. Lett. 77 (1996) 1683. 

\bibitem{sage}
V. Gavrin (SAGE Collaboration) in {\it Neutrino '98} \cite{homestake}. 

\bibitem{gallex}
T. Kirsten (GALLEX Collaboration) in {\em Neutrino '98} \cite{homestake}. 


\bibitem{superkamioka}
Y. Suzuki (SuperKamiokande Collaboration) 
in {\it Neutrino '98} \cite{homestake}. 


\bibitem{JBahcall} J. Bahcall and R. Ulrich,  Rev. of Mod. Phys. {\bf
60}, 297 (1988); J. N. Bahcall, {\it Neutrino Astrophysics}, Cambridge
University Press, New York, (1989).

\bibitem{BP95}
J. N. Bahcall and M.H. Pinsonneault, 
Rev. of Mod. Phys. 67 (1995) 781.

\bibitem{BP98}
J. N. Bahcall, S. Basu and M.H. Pinsonneault, 
Phys. Lett. B 433 (1998) 1. 

\bibitem{reviewSSM}
For a recent review, see J.\ N.\ Bahcall, astro-ph/9808162.

\bibitem{INT}
E. G. Adelberger {\sl et al.}, astro-ph/9805121, 
Rev. Mod. Phys., (to be published, October 1998).

\bibitem{BKS}
J. N. Bahcall, P. I. Krastev and A. Yu. Smirnov,
Phys. Rev. D58 (1998) 096016. 

 
\bibitem{GRC97} M.M. Guzzo, N. Reggiani and J.H. Colonia, Phys. 
Rev. D56 (1997) 588.

\bibitem{priest} E.R.  Priest, {\it Solar
Magnetohydrodynamics}, D.Reidel Publishing Company (1987).

\bibitem{foukal} P.  Foukal, {\it Solar Astrophysics},
Wiley-Interscience Publication (1989).



\bibitem{sno} G.T. Ewan, ``Sudbury Neutrino Observatory", 
in Frontiers
of Neutrino Astrophysics, edited by Y. Suzuki and 
K. Nakamura, Universal
Academy Press, Tokyo (1993) p. 135.



\bibitem{density} P.I. Krastev and A. Yu. Smirnov, 
Phys. Lett. B 226
(1989) 341; Mod. Phys. Lett. A 6 (1991) 1001; 
A. Abada and S.T. Petcov,
Phys. Lett B 279 (1992) 153; F.N. Loretti and 
A.B. Balantekin,
Phys. Rev. D 50 (1994) 4762.

\bibitem{pulido} For a review on neutrino 
spin-flavor precession see J.  
Pulido, Phys.  Rep.  211 (1992) 167.



\bibitem{goedbloed} J.P Goedbloed and P.H. Sakanaka, 
The Physics of
Fluids 17 (1974) 908.

\bibitem{accumulation} J.P. Goedbloed, Physica 12D (1984) 107.

\bibitem{villard} L. Villard. K. Appert, R. Grubber 
and J. Vaclavik, 
Comp. Phys. Rep. 4 (1986) 95.

\bibitem{tataronis} J. M. Kappraf and J. A. Tataronis,
J. Plasma Physics 18 (1977) 209;  T. Sakurai, M. Goossens and 
J. V. Hollweg, Solar Physics 133 (1991) 227.

\bibitem{SR}
J. Schmitt and R. Rosner, Astrophys. J. 265 (1983) 901. 

\bibitem{shi}
X. Shi {\em et al.}, Comm. Nucl. Part. Phys. 
21 (1993) 151. 



\bibitem{hiroshi_npb96} H. Nunokawa, A. Rossi, 
V.B. Semikoz and J.W.F. 
Valle, Nuc. Phys. B472 (1996) 495.

\bibitem{9dohiroshi} S. Turck-Chi\'eze et al., 
Phys. Rep. 230 (1993) 57.

\bibitem{fiorentini} G. Fiorentini and B. Ricci, 
``Solar neutrinos: where 
we are and what is next?'' ASTRO-PH/9801185.


\bibitem{field} Some mechanisms to generate 
such fields are suggested in A. Cisneros
in rerefence [1] and D. Moss, Monthly Notes 
R. Astron. Soc. 224 (1987) 1019.

\bibitem{parker} E. Parker, Astrophys. J. 198 (1975) 205.

\bibitem{bounds} V.D. Kuznetsov and S.I. Syrovastskii, 
Sov. Astron. 23 (1979) 715;
V.N. Krivodubskii, Sov. Astron. Lett. 13 (1987) 338.

\bibitem{somewhere} J.H. Colonia, P.C. de Holanda, 
M.M. Guzzo and N. Reggiani,
in preparation.

\bibitem{footnote} {Obviously, if fluctuations of exact one day
period is observed in the solar neutrino data, MSW 
effect at the Earth
is a natural explanation to be evoked. However any 
fluctuation with
different from exact one day period  rules out MSW.}

\bibitem{GB92} M.M. Guzzo and J. Bellandi, 
Phys. Lett. B317 (1993) 130.




\end{thebibliography}
\end{document}